\documentclass[conference,a4paper]{IEEEtran}

\usepackage[utf8]{inputenc} 
\usepackage[T1]{fontenc}
\usepackage{url} 
\usepackage{cite}  
\usepackage{xcolor}
\usepackage{amssymb,algorithm,algorithmic,amsthm}
\usepackage[cmex10]{amsmath} 
\usepackage{verbatim}

\usepackage{mleftright}       
\mleftright                   

\usepackage{graphicx}
\usepackage{booktabs} 

\def\Ts{T}

\def\calG{{\mathcal{G}}}
\def\calH{{\mathcal{H}}}
\def\calK{{\mathcal{K}}}

\def\calS{{\mathcal{S}}}

\def\b0{{\pmb{0}}}

\def\bg{{\boldsymbol{g}}}

   \def\bV{{\boldsymbol{V}}}
   \def\bW{{\boldsymbol{W}}}
   
   \def\bY{{\boldsymbol{Y}}}

\def\calB{{\mathcal{B}}}
\def\calT{{\mathcal{T}}}
\def\calI{{\mathcal{I}}}

\def\Pr{{\mathsf{P}}}

\def\ua{{\underline{a}}}

\def\Re{{\text{Re}}}
\def\Im{{\text{Im}}}

\newcommand{\argmax}{\mathop{\mathrm{argmax}}}

\newtheorem{theorem}{Theorem}

\newtheorem{proposition}{Proposition}

\newcommand{\dg}[1]{{#1}}

\begin{document}

\IEEEoverridecommandlockouts 

\title{An Asynchronous Massive Access Scheme with Dynamic Range Considerations\thanks{This material is based upon work supported by the National Science Foundation under Grant Nos.~1910168 and~2132700.\\
\indent This is the complete version of the paper, which includes all appendices, to be presented at the IEEE ISIT 2023.}} 

	\author{%
		\IEEEauthorblockN{Lina Liu and Dongning Guo}
		\IEEEauthorblockA{Department of Electrical and
Computer Engineering, Northwestern University, Evanston, Illinois\\
			Email: linaliu2020@u.northwestern.edu; dGuo@northwestern.edu}
	}

	\maketitle
	
\begin{abstract}
    This paper studies the performance of a transmission and reception scheme for massive access under some practical challenges.
    One challenge is the near-far problem, i.e., an access point often receives signals from different transmitting devices at vastly different signal strengths.  Another challenge is that the signals from different devices may be subject to arbitrary, analog, and heterogeneous delays.  This paper considers a fully asynchronous model which is more realistic than the frame or symbol level synchrony needed in most existing work.   
    A main theorem characterizes the asymptotic scaling of the codelength with the number of devices, a device delay upper bound, and the dynamic range of received signal strengths across devices.  The scaling result suggests potential advantages of grouping devices with similar received signal strengths and letting the groups use time sharing.
    The performance of the proposed scheme is evaluated using simulations with and without grouping. 
	\end{abstract}
	
	

	\section{Introduction}
	\label{sec:intro}

	In the 
    context of
    Internet of Things, a \dg{massive} 
    number of sensors and intelligent devices 
    \dg{access the network via an infrastructure of wireless access points (APs)}~\cite{chen2020massive, wu2020massive}.
	\dg{In the absence of power control,}
	signals 
    from 
    \dg{different} devices \dg{may be received at dramatically different strengths (e.g., up to 100 dB)} 
    due to \dg{fading and path loss, including shadowing effects}. 
    \dg{A consequence of such near-far effects is that an AP}
    may fail to detect \dg{or decode} a weak signal from a distant \dg{device in the presence of a strong interfering} 
    signal from a nearby \dg{device.  In addition,} 
    signals transmitted 
    \dg{by devices at different locations generally}
    reach 
    \dg{an AP} with 
    \dg{different} delays.
    \dg{The confounding factors of unknown device delays, amplitues, and phases}
    pose significant challenges to \dg{reliable and efficient massive access}.

The near-far \dg{problem} 
has been extensively studied in code-division multiple-access (CDMA)~\cite{lupas1989linear,  madhow1997blind, muqattash2003solving, torrieri2012guard} and \dg{in} random access mechanisms~\cite{liva2010graph,stefanovic2014exploiting,khaleghi2017near}.
Existing solutions and results are not 
\dg{directly applicable to asynchronous} massive access.
To the best of our knowledge, 
this work presents 
\dg{a} first investigation into 
the near-far problem in the massive access setting.  
\dg{The goal here is to characterize the joint effect of the delay uncertainty and the}
dynamic range of the received \dg{device} signal-to-noise ratios (SNRs).

\dg{In the last decade or so, there have been many studies of}
the scenario where 
only a small fraction of \dg{a very large number of} 
devices 
\dg{transmit in any given slot, each with a very small payload}.
Compressed sensing 
\dg{techniques have been developed for such}
massive connectivity and sporadic traffic~\cite{zhang2013neighbor, liu2018massive}. 
\dg{The role of multiple antennas has also been studied (see, e.g.},~\cite{haghighatshoar2018improved}). Asynchronous \dg{compressed sensing has}
been \dg{proposed}, 
but 
\dg{the} computational complexity \dg{is at least} linear in the device population \cite{applebaum2012asynchronous,liu2021efficient}, 
\dg{and it is difficult to obtain} 
a theoretical performance guarantee \cite{amalladinne2019asynchronous}. Asynchronous covariance-based methods have 
been designed \dg{with few result on how the performance scale with the system size}
~\cite{wang2022covariance,li2022asynchronous}.
\dg{Reference~\cite{shahi2022strongly}
only addresses the scenario of extremely low traffic and extremely high uncertainty about delays.}
	
	


    \dg{In this paper, we consider devices whose SNRs are} within a known dynamic range \dg{and whose delays are analog and constrained by a known upper bound}.
    Building upon our previous work~\cite{chen2022asynchronous}, we explore a signaling scheme based on sparse orthogonal frequency-division multiple-access (OFDMA). This scheme
    \dg{is suitable for the asynchronous setting as the frequency of tones are delay invariant.
    The receiving method in~\cite{chen2022asynchronous}, however, requires the received signals from active devices to have symbol level synchrony (but not frame level synchrony).  Upon receiving a superposition of signals from an unknown subset of devices, the AP performs an iterative algorithm to identify one device at a time, decode the device's message, estimate the device's exact delay, cancel the signal from this device, and then iterate.  The method has only been numerically evaluated in~\cite{chen2022asynchronous} for a very small dynamic range (about 6 dB).
} 

    \dg{Our main contributions are summarized as follows:}
	\begin{itemize}
		\item 
        We modify the sparse OFDMA scheme 
        \dg{of}~\cite{chen2022asynchronous} to \dg{allow arbitrary, analog, and heterogeneous device delays.  In particular, our decoding method and analysis address the challenge of error propagation due to the (analog) delay, amplitude, and phase estimation errors.}
		\item We 
        \dg{characterize a large-system asymptotic} performance of \dg{the scheme for} device identification and message decoding as a function of \dg{system size}, the dynamic range, and a device delay upper bound.

        \item \dg{We show}
        that, when the dynamic range is high, \dg{it is more efficient to} divide 
        the devices into groups of lower dynamic ranges for time sharing.
        \dg{This is validated by simulations under some practical settings.}
        Grouping can be implemented using a beacon signal from the AP, which allows devices to infer about their SNRs in the uplink.
\end{itemize}

 \section{System Model and Scaling Result}\label{sec:system}
	
We study the uplink of a 
\dg{wireless} system with up to $N$ \dg{transmitters or} devices.  Let time be slotted.  In a given slot, 
a random subset 
of devices \dg{transmit.}
We assume that each \dg{of those devices} independently chooses a message from a set of $S$ messages \dg{to send}.
\dg{We} focus on a single \dg{receiver or AP, which needs only to decode messages from devices which are received at sufficiently high amplitude.}
Let $\mathcal{K} \subseteq \{0, \cdots, N-1\}$
denote the set of identities of those devices.

We use OFDM modulation over a frequency band of $B\cdot f$ hertz, where $B\ll N$ stands for the
FFT length and $f$ stands for the bandwidth of each subcarrier.
\dg{Let} $\mathcal{B}_k \subseteq \{ 0, \cdots, B-1 \}$
denote the \dg{sparse subset of} subcarriers assigned to device $k$. 
When active, device $k$ transmits $C$ OFDM symbols, 
\dg{where, in the $c$-th OFDM symbol, the same symbol $g_k^c$ is modulated onto all subcarriers in $\mathcal{B}_k$.  Device $k$'s identity and message are encoded in the subset $\mathcal{B}_k$ and $\big(g_k^0,\dots,g_k^{C-1}\big)$}.

\dg{The duration of} an OFDM symbol 
\dg{is} $1/f$. 
\dg{Let $T=1/(Bf)$.}
We assume that all device delays are 
\dg{positive and upper bounded by $M\Ts$ for some known integer $M\ll B$}.  A cyclic prefix of length $MT$ is \dg{used}. 
The 
{continuous-time baseband} signal transmitted by device $k$ 
in the $c$-th OFDM symbol interval can be expressed as\footnote{The discrete-time chips of each OFDM symbol are obtained by performing an inverse FFT on the frequency-domain sequence and then adding a cyclic prefix.  If the rectangular pulse shaping is used as is common in practice, the resulting transmitted signal differs slightly from the superposition of complex sinusoids
in~\eqref{eq:ofdmsymbol}. Nevertheless, the received signal in the frequency domain can be accurately 
described by~\eqref{eq:Ybc} and~\eqref{eq:Akb}.}
 \begin{align}\label{eq:ofdmsymbol}
    &s_k^c(t) = g_{k}^c \sum_{b \in \mathcal{B}_k} e^{\iota 2\pi b 
    f (t-c(B+M)T)} .
\end{align}
In Secs.~\ref{sec:proof} and~\ref{sec:simulation}, we show a design where each device uses as few as three subcarriers, in which case the superposition~\eqref{eq:ofdmsymbol} may be generated directly without using FFT.

    \dg{Assuming constant} channel coefficient $a_k\in\mathbb{C}$ and 
    delay $\tau_k$ \dg{from device $k$ to the AP over the entire transmission slot, we express} the 
    \dg{corresponding} received signal 
    \dg{as}
	\begin{align}\label{eq:system_model}
	 a_k s_k^c(t-\tau_k) 
	&= a_{k} g_{k}^c \sum_{b \in \mathcal{B}_k} e^{-\iota 2\pi bf\tau_k} e^{\iota 2\pi bf (t-c(B+M)T)}. 
	\end{align}
    \dg{Importantly, the impact of the delay $\tau_k$ on subcarrier $b$ is exactly}
    a 
    phase shift of
	$2\pi b f\tau_k$.
	
	
	The \dg{AP receives} 
    a superposition of 
    device signals \dg{which is corrupted by} 
    an additive 
    white Gaussian noise 
    with double-sided power spectral density of $\sigma^2$.
    \dg{Following~\cite{chen2022asynchronous}, we express}
    the received signal on subcarrier $b$ \dg{(in the frequency domain) in} the $c$-th OFDM symbol \dg{as} 
   \begin{align}
 Y^c_b 
	&= \sum_{k \in \mathcal{K}: b \in \mathcal{B}_k} A_{k,b} g^c_{k} + W^c_b , \quad b = 0, \cdots, B-1
    ,\label{eq:Ybc}
\end{align}
	where 
	\begin{align}\label{eq:Akb}
	A_{k,b}=a_ke^{-\iota 2\pi b \tau_k/(B\Ts )},
	\end{align}
	and $W^c_b$ are i.i.d.\ \dg{circularly symmetric} complex Gaussian variables with 
    \dg{variance $\sigma^2/B$ per dimension, where the factor of $B$ is due to  normalization of the device signal.}

\begin{theorem}\label{thm:asyn}
    Suppose out of $N$ devices, each with a message set containing no more than $S$ messages, $K$ unknown devices turn active and transmit to the AP \dg{in a given slot}. Each active device sends one message with a delay no more than 
    \dg{$M$ samples.  Let $\ua$ and $\bar{a}$ denote a lower bound and an upper bound, respectively, for the received signal amplitudes of those devices.}
    Then there exist constant \dg{integers} $\beta_0,\beta_1$ and $\beta_2$ \dg{such that} for any $\ua,\bar{a},\epsilon>0$ with $\ua<\bar{a}$, 
    \dg{there exists} $K_0$ such that for every $N\geq K\geq K_0$ and every natural number $S$, all \dg{those} active devices' identities and messages can be decoded correctly with probability no less than $1-\epsilon$ using OFDMA with a codelength \dg{of}  
	\begin{align}\label{eq:L}
    \begin{split}
		L&=(\beta_0K+M)(\lceil\lceil \log(NS) \rceil/R\rceil+1+\lceil\beta_1\log K\rceil)\\
		&\qquad+M+\left\lceil\beta_2(\bar{a}/\ua)^2(\log K)^4K\log (KM+1)\right\rceil .
    \end{split}
	\end{align}
	\end{theorem}

    \dg{In a typical massive access system, the delay bound $M$ is a small constant and a very small fraction of devices transmit near a given AP, hence the theorem implies that the codelength is polynomial in $K$ and sublinear in the device population $N$.}

    \dg{Moreover,} $\bar{a}/\ua$ captures the dynamic range of the received 
    \dg{amplitudes},
    which is a significant \dg{factor in~\eqref{eq:L}. In particular, if the first additive term therein is not dominant,} 
    the codelength can be reduced substantially by dividing devices into groups with smaller dynamic range for \dg{``time sharing'' of each time slot, as will also} 
    be demonstrated \dg{numerically} in Sec.~\ref{sec:simulation}. 

\section{Sparse OFDM Signaling \dg{Design}}
\label{sec:signaling}


\dg{We let all devices use the same number ($D$) subcarriers.  In particular,}
$\calB_k$ is randomly chosen among all possible $D$-element subsets of the subcarriers. 
\dg{In a given time slot,} a subcarrier is referred to as a \textit{zeroton}, \textit{singleton}, or \textit{multiton}, if no device, a single device, or multiple devices transmit on the subcarrier, respectively.

\dg{In a given time slot for an active device, we let the device transmit three subframes in tandem.  The remainder of this section describes the design of those subframes (0, 1, and 2).}
	
\subsection{\dg{Subframes 0}}
\label{sec:sig_comp1}

    \dg{Subframe 0 consists of $C_0$ OFDM symbols used} to encode the device identity \dg{and message.  Together, the identity bits and message bits are treated as $\lceil \log (NS) \rceil$ information bits.  To combat noise and other uncertainty, these information bits are encoded using a rate $R$ code into $C_0=1+\lceil\lceil\log(NS)\rceil/R\rceil$ BPSK symbols, denoted as $\tilde{\bg}_{k} \in 
    \{-1,+1\}^{C_0}$,
    where the first BPSK symbol is set to $\tilde{g}_k^0=+1$ as a reference for phase recovery (as in~\cite{chen2022asynchronous}). 
    Each of the $C_0$ BPSK symbols is modulated onto all active subcarriers of the corresponding OFDM symbol.

\subsection{\dg{Subframes 1}}

    Subframe 1 consists of} $C_1$ OFDM symbols used to mitigate false alarms.
    \dg{The corresponding BPSK sequence is denoted as $\dot{\bg}_{k}$, 
    whose entries}
     are generated according to  
     $\mathsf{P} \{\dot{g}_k^c = \pm 1 \} = 1/2$, $c=0, \cdots, C_1-1$.

%
Denote by $\bg_k = \left(
	\tilde{\bg}_k^T,\dot{\bg}_k^T\right)^T$.
Using~\eqref{eq:Ybc}, we collect 
the $b$-th subcarrier signals over all OFDM symbols as a vector:
\begin{align} \label{eq:yb}
	\bY_b &= \left(
	\begin{array}{c}
	\tilde{\bY}_b \\
	\dot{\bY}_b 
	\end{array}
	\right) \\
	\label{eq:DFT}& =\sum_{k \in \mathcal{K}: b \in \mathcal{B}_k } A_{k,b}  \left(
	\begin{array}{c}
	\tilde{\bg}_{k} \\
	\dot{\bg}_{k} 
	\end{array}
	\right)
	+
	\left(
	\begin{array}{c}
	\tilde{\bW}_b \\
	\dot{\bW}_b 
	\end{array}
	\right)
	\end{align}
	where the dimensions of $\tilde{\bY}_b$ and $\dot{\bY}_b$ are $C_0$ and $C_1$, respectively, so are the dimensions of  $\tilde{\bW}_b$ and $\dot{\bW}_b$.

	\subsection{\dg{Subframe 2}}
    \label{sec:sig_comp2}
    \dg{Subframe 2} 
    consists of 
    $C_2$ (time-domain) BPSK chips, each with a duration of ${\Ts }$.
    Once a device's identity is decoded from subframes 0 and 1, the AP uses the device's subframe 2 to estimate the device delay so as to reconstruct the signal for cancellation.

	\section{Device Identification and Message Decoding}
	

    We first briefly describe the scheme    in~\cite{chen2022asynchronous} for classifying subcarriers as zeroton, singleton, or multiton, and 
    also for decoding using singleton subcarriers.
    After introducing a new delay estimation scheme 
    to aid in cancellation 
    of the decoded devices' signals,
    we describe the overall identification and decoding scheme.
	
	\subsection{Robust Subcarrier Detection}\label{sec:subcarrier_detection}

	
	\subsubsection{Zeroton detection}
	
	Subcarrier $b$ is declared to be a zeroton if $\Vert\dot{\bY}_b\Vert^2<\eta$, where $\Vert\cdot\Vert$ denote the $\ell^2$-norm, and $\eta$ is some constant threshold.
	
	\subsubsection{Phase estimation}

    Recall that $\tilde{g}_k^0=+1$ is modulated onto all subcarriers assigned to device $k$.
	If subcarrier $b$ is not declared to be a zeroton, the phase of $A_{k,b}$ is estimated as
	\begin{align}\label{eq:phase}
	\hat{\theta}_b&= \angle \tilde{Y}_{b}^0.
	\end{align}
	
	\subsubsection{Device identification and message decoding}
	\label{sec:device_identification}
	
	Using the 
    phase estimate, 
    we perform hard binary decisions on $ \text{Re} \left\{ \tilde{Y}^c_b e^{- \iota \hat{\theta}_b} \right\}$ for $c=1,\cdots,C_0-1$, assuming subcarrier $b$ is a singleton, and then decode the device index and message accordingly.
	\dg{Let the decoded index be denoted as $\hat{k}$.}
 
	\subsubsection{Singleton verification}
	
	Subcarrier $b$ is declared to be a singleton \dg{associated with device $\hat{k}$} if and only if it \dg{satisfies}
	\begin{align}\label{eq:multiton_check}
	\Vert \dot{\bY}_b - \dot{A}_{\hat{k},b} \dot{\bg}_{\hat{k}} \Vert^2 \leq  \eta,
	\end{align}
	where $\dot{A}_{\hat{k}, b} = \frac{1}{C_1} \dot{\bg}_{\hat{k}}^{\dagger} \dot{\bY}_b$ is the estimated non-zero signal. Otherwise, subcarrier $b$ will be declared as a multiton.
	
	
\subsection{Delay Estimation}\label{sec:delay_estimation}
In order to effectively cancel a decoded device's signal, \dg{the device's amplitude and delay need to be estimated sufficiently accurately.}

	
	
	The first $M$ 
    \dg{samples in subframe 2} 
    is discarded to account for the unknown but bounded propagation delay. The remaining signal is collected within the interval
    \begin{align}\label{eq:calI}
	\calI=[C(B+M){\Ts }+M{\Ts },C(B+M){\Ts }+C_2{\Ts })
	\end{align}
    \dg{which contains}
    $I=C_2-M$ \dg{samples}.
    Let $s'_k(t)$ 
    represent the signal transmitted by device $k$ in 
    subrame 2.

    The AP's received signal \dg{in subframe 2 sans noise can be expressed as} 
	\begin{align}
	x'(t)=\sum_{k\in\mathcal{K}}a_k{s'_k}(t-\tau_k),\quad t\in\calI.
	\end{align}
    We focus on a decoded device $k$ and 
    \dg{define the following statistic for delay estimation:}
    \begin{align}\label{eq:Tk}
	\calT_k(\tau)=\int_{t\in\calI}x'(t){s'}_k(t-\tau)dt+\sum_{i=0}^{I-1}Z_i,
	\end{align}
   where $Z_i\sim\mathcal{CN}(0,2\sigma^2{\Ts })$ represents the white Gaussian noise in the $i$-th of the $I$ sampling intervals. 
 It can be calculated that
	\begin{align}\label{eq:Tk_int}
	&\mathcal{T}_k(\tau)=\\
	&\begin{cases}
	a_k\Ts I+a_k|\tau-\tau_k|\left(\sum_{i=0}^{I-1}R_{i,k}-I\right)+\sum_{i=0}^{I-1}Z_i\nonumber\\
	\quad+\sum_{p\in\mathcal{K}\backslash k}a_p\left(\Delta_p\sum_{i=0}^{I-1}R_{i,p}+({\Ts }-\Delta_p)\sum_{i=0}^{I-1}{R'}_{i,p}\right)\nonumber\\
	\qquad\qquad\qquad\qquad\qquad\qquad\qquad\text{if }|\tau-\tau_k|\leq {\Ts },\\
	\sum_{k\in\mathcal{K}}a_k\left(\Delta_k\sum_{i=0}^{I-1}R_{i,k}+({\Ts }-\Delta_k)\sum_{i=0}^{I-1}{R'}_{i,k}\right)\nonumber\\
    \quad+\sum_{i=0}^{I-1}Z_i, 
    \qquad\qquad\qquad\text{if }{\Ts }\leq|\tau-\tau_k|\leq  M\Ts ,
	\end{cases}
	\end{align}
	where $R_{i,k}$ and ${R'}_{i,k}$ are i.i.d.~BPSK symbols, and $\Delta_k=(|\tau-\tau_k|\mod {\Ts })$.
 The delay of device $k$ can be estimated as
	\begin{align}\label{eq:est_tau}
	\hat{\tau}_k=\argmax_{0\leq \tau\leq  M\Ts }|\calT_k(\tau)|.
	\end{align}
	
   
    
    To solve \eqref{eq:est_tau} with a desired accuracy, we propose a two-step delay estimation.
	
	\subsubsection{Crude estimation}
	First divide the feasible region $[0,M{\Ts }]$ into slots of length ${\Ts }$. The slot boundaries are $\calH=\{0,{\Ts },2{\Ts },\cdots, M{\Ts }\}$. For each $\tau\in\calH$, calculate the statistic $|\calT_k(\tau)|$. Define the crude estimation threshold as
	\begin{align}\label{eq:Tthre}
	\bar{\calT}=\ua{\Ts } I/4.
	\end{align} 
    Then we declare
	\begin{align}\label{eq:t_crude}
	\tau_k\in\begin{cases}
		[0,{\Ts }],	\quad\text{if only $|\calT_k(0)|$ exceeds $\bar{\calT}$;}\\
		[(M-1){\Ts },M{\Ts }],\quad\text{if  only $|\calT_k(M{\Ts })|$ exceeds $\bar{\calT}$;}\\
		[i{\Ts },(i+1){\Ts }], \quad\text{if only $|\calT_k(i{\Ts })|$ and }|\calT_k((i+1){\Ts })|\\ 
		\qquad\text{ exceed }\bar{\calT}\text{ with some }i\in\{1,\cdots,M-1\};\\
		[i{\Ts }-{\Ts }/2,i{\Ts }+{\Ts }/2],\quad\text{if only $|\calT_k(i{\Ts })|$ exceeds $\bar{\calT}$}\\
		\qquad\text{with some }i\in\{1,\cdots,M-1\}.
	\end{cases}
    \end{align}
    If no condition in~\eqref{eq:t_crude} is satisfied, we declare a 
    delay estimation \dg{failure.  We regard the entire time slot as an error.}
	
	\subsubsection{Refined estimation} 
	Suppose we find a slot in the crude estimation. We further discretize it into finer slots of length
	\begin{align}\label{eq:fine_length}
	\psi = {{\Ts }}/{\left\lceil 2{\Ts }(\log K)^2\big/\rho\right\rceil},
	\end{align}
	so that $\frac{\rho}{4(\log K)^2}\leq \psi\leq \frac{\rho}{2(\log K)^2}$, where $\rho$ is a constant. Represent the obtained slot from the crude estimation as $\lambda+[0,{\Ts }]$, where $\lambda$ stands for the left boundary of the slot. Subtracting the offset $\lambda$, the boundary points of the finer slots can be represented as $\calH'=\left\{0,\psi,2\psi,\cdots,{\Ts }\right\}$. For each $\tau\in\calH'$, calculate the statistic $|\calT_k(\tau+\lambda)|$, and let $\tau^*=\argmax_{\tau\in\calH'}|\calT_k(\tau+\lambda)|$. Then we declare that 
	\begin{align}\label{eq:t_fine}
	&\tau_k\in\left[\max\left\{0,\tau^*-\frac{\rho}{2(\log K)^2}+\lambda\right\},\right.\nonumber\\
	&\qquad\left.\min\left\{\tau^*+\frac{\rho}{2(\log K)^2}+\lambda,M{\Ts }\right\}\right].
	\end{align}
	If the declaration is correct, for any $\hat{\tau}_k$ in the slot given in \eqref{eq:t_fine}, $|\hat{\tau}_k-\tau_k|\leq \frac{\rho}{(\log K)^2}$ .


	\subsection{Overall Identification and Decoding Scheme}\label{sec:overall}
	If the delay estimation process is successful, take any $\hat{\tau}_k$ in the slot given in \eqref{eq:t_fine}. The channel coefficient $a_k$, which is a deterministic parameter, is then estimated according to~\eqref{eq:Akb} as
	\begin{align}\label{eq:channel_estimate}
	\hat{a}_{k} = \frac{1}{C} \bg_k^{\dagger} \bY_{b} e^{\iota 2 \pi b \hat{\tau}_k/(B\Ts ) }.
	\end{align}
	The received signal of a connected unprocessed subcarrier $b'$ is then updated as
	\begin{align}\label{eq:succ_canel}
	\bY_{b'} \leftarrow \bY_{b'} -  \hat{a}_{k} e^{-\iota 2 \pi b' \hat{\tau}_{k}/(B\Ts )} \bg_{k}.
	\end{align}
	On all these subcarriers, the robust subcarrier detection will be implemented, and new zerotons or singletons may arise. Once new devices are detected and decoded on singletons, their information is subtracted from connected subcarriers.  This process of successive cancellation continues until there are no more subcarriers declared as singletons.

	\section{Proof of Theorem \ref{thm:asyn}}\label{sec:proof}
	
The proof extends the work in \cite{chen2022asynchronous} by incorporating the new delay estimation analysis, addressing the error propagation caused by the analog delay, and taking into account the dynamic range of the received device SNRs.

The codeword can be constructed using the following parameters:
	\begin{align}
	& D \geq 3,\label{eq:para_T}\\
	& \beta_0 \geq D(D-1)+1,\\
	& B = \beta_0 K,\label{eq:para_B}\\
	& C_0 = \lceil\lceil \log(NS) \rceil/R\rceil+1,\label{eq:C0}\\
	& C_1 = \lceil\beta_1\log K\rceil,\label{eq:para_C1}\\
	& C_2 = 
	M + \lceil \beta_2(\bar{a}/\ua)^2(\log K)^4K\log(KM+1)\rceil,\label{eq:para_C2}
	\end{align}
	where $\beta_1,\beta_2>0$ and $R<1$ are constants. 
	Thus, $L=(B+M)(C_0+C_1)+C_2$ satisfies \eqref{eq:L}.
	
	
	Let $G$ denote the bipartite graph 
    with $K$ left nodes and $B$ right nodes, 
    which correspond to $K$ active devices and $B$ subcarriers, respectively. The $k$-th left node is connected to the $b$-th right node if device $k$ transmits on subcarrier $b$. 
	When no two active devices share the same set of subcarriers, $G$ is convertible to an equivalent hypergraph, denoted as $G\in\hat{\calG}$~\cite{karonski2002phase}. It is shown that
	\begin{align}\label{eq:hypergraph}
	\mathsf{P}\left\{G\in\hat{\calG}\right\}\geq 1-\zeta/K,
	\end{align}
	with some constant $\zeta$ for sufficiently large $K$.
	
	Let $\calG$ denote the ensemble of $D$-uniform hypergraphs with $K$ hyperedges and $B$ vertices, made up of components that are either a hypertree or a unicyclic component, and none of which have more than $\alpha D\log(\beta_0 K)/(D-1)$ hyperedges with constant $\alpha>0$.
    The following two propositions have been proved in~\cite{chen2022asynchronous}.
	\begin{proposition}\label{prop:prop1}
		Under \eqref{eq:para_T}-\eqref{eq:para_C1}, there exist constants $K_{0}^{\text{h}}>0$ and $\nu>1$, such that for every $K\geq K_{0}^{\text{h}}$, we have
		\begin{align}\label{eq:ensemble}
		\mathsf{P}\left\{G\in\calG|G\in\hat{\calG}\right\}\geq 1-\nu/K.
		\end{align}
	\end{proposition}
	
	\begin{proposition}\label{prop:prop2}
		If $G\in\calG$, all active devices can be correctly detected and decoded, as long as the robust subcarrier detection is always correct during the identification and decoding process.
	\end{proposition}
	
	Let $\calS(t-1)$ be the set of recovered devices in the previous $t-1$ iterations during the identification and decoding process. For each device $\ell\in\calS(t-1)$ decoded from a singleton bin $b_{\ell}$, define the error due to noise as
	\begin{align}\label{eq:e_k}
	e_{\ell}=C^{-1}\bg_{\ell}^{\dagger}\bW_{b_{\ell}}.
	\end{align}
	Clearly, $e_{\ell}\sim\mathcal{CN}(0,2\sigma^2/(BC))$. In iteration $t$, the received signal on subcarrier $b$ can be expressed as
	\begin{align}\label{eq:Yb}
	\bY_b=\sum_{k\in\calK\backslash\calS(t-1):b\in\calB_k}A_{k,b}\bg_k+\bW_b+\bV_b,
	\end{align}
	where the sum is taken over all active devices that are hashed to subcarrier $b$ and have not yet been recovered, and $\bV_b$ is caused by residual channel estimation errors and delay estimation errors from 
    previously recovered devices.
	
	\begin{proposition}\label{prop:prop3}
		Suppose \eqref{eq:para_T}-\eqref{eq:para_C2} hold and $G\in\calG$.
		Suppose the robust subcarrier detection is always accurate in detecting every $\ell\in\mathcal{S}(t-1)$. Suppose the amplitude of the error due to noise is upper bounded by
		\begin{align}\label{eq:ub_ek}
		|e_{\ell}|\leq\varrho(\log K)^{-2},
		\end{align}
		where
		\begin{align}\label{eq:ub_varrho}
		\varrho=\frac{(D-1)\sqrt{\eta}}{8\alpha\beta_1(1+\log\beta_0)D},
		\end{align}
		and the delay estimation error is upper bounded by
		\begin{align}\label{eq:ub_epsilonk}
			|\epsilon_{\ell}|\leq \rho(\log K)^{-2},
		\end{align}
		where
		\begin{align}\label{eq:ub_rho}
			\rho = \frac{{\Ts }(D-1)\sqrt{\eta}}{32\pi \bar{a}\alpha\beta_1(1+\log\beta_0)D}.
		\end{align} 
		Then, there exits a constant $K_{0}^{\text{r}}$, such that for every $K\geq K_{0}^{\text{r}}$, every entry of $\bV_b$ is bounded by 
		\begin{align}\label{eq:ub_vbc}
		|V_b^c|\leq \frac{\sqrt{\eta}}{\beta_1\log K}.
		\end{align}
	\end{proposition}

    Proposition~\ref{prop:prop3} is proved in Appendix~\ref{app:prop3}.
	
	\begin{proposition}\label{prop:prop4}
		Suppose \eqref{eq:para_T}-\eqref{eq:para_C2} hold and $G\in\calG$. For $K\geq K_0^c$ with $K_0^c$ being a sufficiently large constant, we can choose $\eta=\ua^2$ such that for every $t=1,2,\cdots$, conditioned on that the robust subcarrier detection is always correct in the first $t-1$ iterations during the identification and decoding process, $e_{\ell}$ is upper bounded by \eqref{eq:ub_ek} and $\epsilon_{\ell}$ is upper bounded by \eqref{eq:ub_epsilonk} for each decoded device $\ell$ in the first $t-1$ iterations, the robust subcarrier detection makes a wrong decision with probability no greater than $7K^{-2}$ in the $t$-th iteration. Moreover, if device $k$ is decoded from a singleton bin $b_k$ in the $t$-th iteration, then $|e_k|\geq\varrho(\log K)^{-2}$ with probability smaller than $K^{-2}$, and $|\epsilon_k|\geq\rho(\log K)^{-2}$ with probability smaller than $(28+128D/(\rho M))K^{-2}$. 
	\end{proposition}
	
    Proposition~\ref{prop:prop4} is proved in Appendix \ref{app:prop4}. Assuming Propositions~\ref{prop:prop1}-\ref{prop:prop4} hold, we upper bound the massive access error probability $P_s$ as follows. Let $\mathcal{E}$ denote the event that the robust subcarrier detection makes at least one wrong decision during the successive identification and decoding process. By Proposition~\ref{prop:prop2}, massive access succeeds if $G \in \mathcal{G}$ and that $\mathcal{E}$ does not occur. Evidently,
	\begin{align}
	P_s&\leq\mathsf{P} \{\mathcal{E}\cup (G\notin\mathcal{G})\}\\
	&\leq \mathsf{P} \{\mathcal{E}\cap (G\in\mathcal{G})\}+\mathsf{P}\{G\notin\mathcal{G}\}\\
	\label{eq:supp_fail}&\leq\mathsf{P} \{ \mathcal{E} | G \in \mathcal{G}\} + \mathsf{P} \{ G \notin \mathcal{G}|G\in\hat{\mathcal{G}}  \}+\mathsf{P}\{G\notin\hat{\mathcal{G}}\}.
	\end{align}
	Every time a device is recovered, the robust subcarrier detection is performed on its connected subcarriers. Since there are $K$ active devices and each of them is connected to $D$ subcarriers, it runs for at most $K D$ times throughout the detection process. By the union bound and the result of Proposition~\ref{prop:prop4},
	\begin{align}
	\mathsf{P} \{\mathcal{E} | G \in \mathcal{G} \} 
	&\leq KD\left(\frac{7}{K^2}+\frac{1}{K^2}+\frac{28+128D}{\rho M}\frac{1}{K^2}\right)\\
	& =\left(8+\frac{28+128D}{\rho M}\right)\frac{D}{K}\label{eq:union_bin_error}.
	\end{align}
	
	Plugging \eqref{eq:hypergraph}, \eqref{eq:ensemble}, and \eqref{eq:union_bin_error} into \eqref{eq:supp_fail}, we have that massive access fails with probability 
	\begin{align}
	P_s \leq \left(\nu+\zeta+8D+\frac{28D+128D^2}{\rho M}\right)\frac{1}{K}.
	\end{align}
	Therefore, given the choice of $D$, $B$, and $C$, massive access fails with probability less than $\epsilon$ as long as $K \geq \max\left\{\left(\nu+\zeta+8D+\frac{28D+128D^2}{\rho M}\right)/\epsilon,K_0^{\text{h}},K_0^{\text{r}},K_0^{\text{c}}\right\}$.

	\section{Simulation Results}\label{sec:simulation}
    In our setting, the message bits and identity bits are interchangeable.
    It suffices to only consider device identification and set $S=1$ in simulations. Assume each device's distance to the AP is uniformly distributed. The channel coefficient can be modeled as $Gd^{-\alpha}$, where $d$ is the device's distance to the AP, $\alpha=3$ is chosen as the path loss exponent, and $G$ is the Rayleigh fading. With fixed $\sigma^2$, $\bar{a}$ and $\ua$ are set to achieve different SNR dynamic ranges, i.e., $20\log_{10}(\bar{a}/\ua)$ dB. Amplitudes outside the range of $(\ua,\bar{a})$ will be regarded as outage. The lowest received SNR is defined as 
    $10\log_{10}(\ua^2/2\sigma^2)$. Following parameters are set throughout the simulations: the total number of devices $N=2^{38}$ is very large; the FFT length $B=6 K$, the delay upper bound $M=20$, $C_0=2+2\lceil\log N\rceil$, $C_1=\lceil\log K\rceil$, and $D=3$. 

	\begin{figure}
		\centering
		\includegraphics[width=0.9\columnwidth]{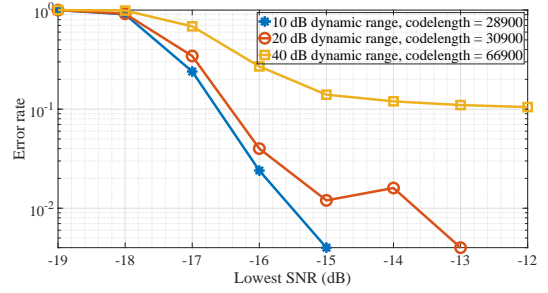}
		\caption{With the increase of the lowest SNR, the error rate of the massive access approaches 0.}
		\label{fig:dynamic}
	\end{figure}

    We first choose $K=50$. We consider three dynamic ranges, 10 dB, 20 dB, and 40 dB, which are associated with $C_2=2000$, 4000, and 40000, respectively. The error rates of the device identification are plotted in Fig.~\ref{fig:dynamic}.
    We observe that when the dynamic range is large, the codelength (given in the legend) increases significantly.  In this case, we can divide the active devices into two groups with a smaller dynamic range and transmit their data in separate time frames. The resulting codelength is the sum of the new codelengths of the two groups.  An error occurs when device identification fails in either group or both. 
	
\begin{figure}
		\centering
		\includegraphics[width=0.9\columnwidth]{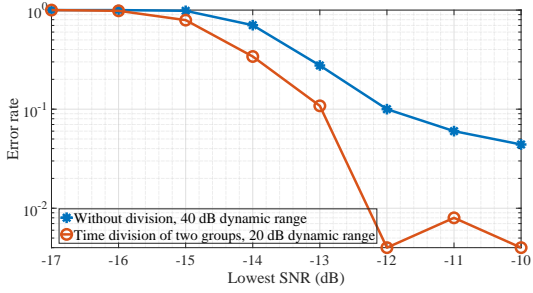}
		\caption{By dividing active devices into two groups, the codelength can be reduced, and the error rate is also reduced.}
		\label{fig:divide}
	\end{figure}

    For a dynamic range of 40 dB with $K=20$ active devices, we divided the devices into two groups each with a dynamic range of 20 dB. We set $C_2=20000$ without division, and $C_2=3000$ with division. Depending on the division, the maximum codelength with time division is 29000, which is smaller than the codelength of 31640 without division. The error rates of the device identification are plotted in Fig. \ref{fig:divide}.
    With grouping and time division, it is possible to achieve a smaller error rate with a shorter codelength.
 
	\section{Conclusion}
	\label{sec:conclusion}
	
    We have \dg{adapted and} analyzed a 
    sparse OFDMA scheme for asynchronous massive access with arbitrary continuous delays \dg{and potentially strong near-far effects}. 
    \dg{The theoretical bound and numerical results provide some guidance on how to}
    effectively manage 
    the \dg{near-far effect in practice}.

	
	\bibliographystyle{IEEEtran}
	\bibliography{isit_bib}

\begin{thebibliography}{10}
\providecommand{\url}[1]{#1}
\csname url@samestyle\endcsname
\providecommand{\newblock}{\relax}
\providecommand{\bibinfo}[2]{#2}
\providecommand{\BIBentrySTDinterwordspacing}{\spaceskip=0pt\relax}
\providecommand{\BIBentryALTinterwordstretchfactor}{4}
\providecommand{\BIBentryALTinterwordspacing}{\spaceskip=\fontdimen2\font plus
\BIBentryALTinterwordstretchfactor\fontdimen3\font minus
  \fontdimen4\font\relax}
\providecommand{\BIBforeignlanguage}[2]{{%
\expandafter\ifx\csname l@#1\endcsname\relax
\typeout{** WARNING: IEEEtran.bst: No hyphenation pattern has been}%
\typeout{** loaded for the language `#1'. Using the pattern for}%
\typeout{** the default language instead.}%
\else
\language=\csname l@#1\endcsname
\fi
#2}}
\providecommand{\BIBdecl}{\relax}
\BIBdecl

\bibitem{chen2020massive}
X.~Chen, D.~W.~K. Ng, W.~Yu, E.~G. Larsson, N.~Al-Dhahir, and R.~Schober,
  ``Massive access for 5{G} and beyond,'' \emph{IEEE Journal on Selected Areas
  in Communications}, vol.~39, no.~3, pp. 615--637, 2020.

\bibitem{wu2020massive}
Y.~Wu, X.~Gao, S.~Zhou, W.~Yang, Y.~Polyanskiy, and G.~Caire, ``Massive access
  for future wireless communication systems,'' \emph{IEEE Wireless
  Communications}, vol.~27, no.~4, pp. 148--156, 2020.

\bibitem{lupas1989linear}
R.~Lupas and S.~Verdu, ``Linear multiuser detectors for synchronous
  code-division multiple-access channels,'' \emph{IEEE transactions on
  Information Theory}, vol.~35, no.~1, pp. 123--136, 1989.

\bibitem{madhow1997blind}
U.~Madhow, ``Blind adaptive interference suppression for the near-far resistant
  acquisition and demodulation of direct-sequence {CDMA} signals,'' \emph{IEEE
  Transactions on Signal Processing}, vol.~45, no.~1, pp. 124--136, 1997.

\bibitem{muqattash2003solving}
A.~Muqattash, M.~Krunz, and W.~E. Ryan, ``Solving the near--far problem in
  {CDMA}-based ad hoc networks,'' \emph{Ad Hoc Networks}, vol.~1, no.~4, pp.
  435--453, 2003.

\bibitem{torrieri2012guard}
D.~Torrieri and M.~C. Valenti, ``Guard zones and the near-far problem in
  {DS-CDMA} ad hoc networks,'' in \emph{MILCOM 2012-2012 IEEE Military
  Communications Conference}.\hskip 1em plus 0.5em minus 0.4em\relax IEEE,
  2012, pp. 1--6.

\bibitem{liva2010graph}
G.~Liva, ``Graph-based analysis and optimization of contention resolution
  diversity slotted {ALOHA},'' \emph{IEEE Transactions on Communications},
  vol.~59, no.~2, pp. 477--487, 2010.

\bibitem{stefanovic2014exploiting}
{\v{C}}.~Stefanovi{\'c}, M.~Momoda, and P.~Popovski, ``Exploiting capture
  effect in frameless {ALOHA} for massive wireless random access,'' in
  \emph{2014 IEEE Wireless Communications and Networking Conference
  (WCNC)}.\hskip 1em plus 0.5em minus 0.4em\relax IEEE, 2014, pp. 1762--1767.

\bibitem{khaleghi2017near}
E.~E. Khaleghi, C.~Adjih, A.~Alloum, and P.~M{\"u}hlethaler, ``Near-far effect
  on coded slotted {ALOHA},'' in \emph{2017 IEEE 28th Annual International
  Symposium on Personal, Indoor, and Mobile Radio Communications
  (PIMRC)}.\hskip 1em plus 0.5em minus 0.4em\relax IEEE, 2017, pp. 1--7.

\bibitem{zhang2013neighbor}
L.~Zhang, J.~Luo, and D.~Guo, ``Neighbor discovery for wireless networks via
  compressed sensing,'' \emph{Performance Evaluation}, vol.~70, no. 7-8, pp.
  457--471, 2013.

\bibitem{liu2018massive}
L.~Liu and W.~Yu, ``Massive connectivity with massive {MIMO}—{P}art i: Device
  activity detection and channel estimation,'' \emph{IEEE Transactions on
  Signal Processing}, vol.~66, no.~11, pp. 2933--2946, 2018.

\bibitem{haghighatshoar2018improved}
S.~Haghighatshoar, P.~Jung, and G.~Caire, ``Improved scaling law for activity
  detection in massive {MIMO} systems,'' in \emph{2018 IEEE International
  Symposium on Information Theory (ISIT)}.\hskip 1em plus 0.5em minus
  0.4em\relax IEEE, 2018, pp. 381--385.

\bibitem{applebaum2012asynchronous}
L.~Applebaum, W.~U. Bajwa, M.~F. Duarte, and R.~Calderbank, ``Asynchronous
  code-division random access using convex optimization,'' \emph{Physical
  Communication}, vol.~5, no.~2, pp. 129--147, 2012.

\bibitem{liu2021efficient}
L.~Liu and Y.-F. Liu, ``An efficient algorithm for device detection and channel
  estimation in asynchronous {I}o{T} systems,'' in \emph{ICASSP 2021-2021 IEEE
  International Conference on Acoustics, Speech and Signal Processing
  (ICASSP)}.\hskip 1em plus 0.5em minus 0.4em\relax IEEE, 2021, pp. 4815--4819.

\bibitem{amalladinne2019asynchronous}
V.~K. Amalladinne, K.~R. Narayanan, J.-F. Chamberland, and D.~Guo,
  ``Asynchronous neighbor discovery using coupled compressive sensing,'' in
  \emph{ICASSP 2019-2019 IEEE International Conference on Acoustics, Speech and
  Signal Processing (ICASSP)}.\hskip 1em plus 0.5em minus 0.4em\relax IEEE,
  2019, pp. 4569--4573.

\bibitem{wang2022covariance}
Z.~Wang, Y.-F. Liu, and L.~Liu, ``Covariance-based joint device activity and
  delay detection in asynchronous m{MTC},'' \emph{IEEE Signal Processing
  Letters}, vol.~29, pp. 538--542, 2022.

\bibitem{li2022asynchronous}
Y.~Li, Q.~Lin, Y.-F. Liu, B.~Ai, and Y.-C. Wu, ``Asynchronous activity
  detection for cell-free massive{ MIMO}: From centralized to distributed
  algorithms,'' \emph{IEEE Transactions on Wireless Communications}, 2022.

\bibitem{shahi2022strongly}
S.~Shahi, D.~Tuninetti, and N.~Devroye, ``The strongly asynchronous massive
  access channel,'' \emph{Entropy}, vol.~25, no.~1, p.~65, 2022.

\bibitem{chen2022asynchronous}
X.~Chen, L.~Liu, D.~Guo, and G.~W. Wornell, ``Asynchronous massive access and
  neighbor discovery using {OFDMA},'' \emph{IEEE Transactions on Information
  Theory}, 2022.

\bibitem{karonski2002phase}
M.~Karo{\'n}ski and T.~{\L}uczak, ``The phase transition in a random
  hypergraph,'' \emph{Journal of Computational and Applied Mathematics}, vol.
  142, no.~1, pp. 125--135, 2002.

\bibitem{rivasplata2012subgaussian}
O.~Rivasplata, ``Subgaussian random variables: An expository note,''
  \emph{Internet publication, PDF}, vol.~5, 2012.

\end{thebibliography}
	
	%
	%
	%
	%
	%
	%
	%
	%
	%
	
	\appendix
	
\subsection{Proof of Proposition \ref{prop:prop3}}\label{app:prop3}
	Consider the first iteration and a singleton subcarrier $b_k$ due to device $k$. The frequency value of subcarrier-$b_k$ is given by
	\begin{align}
	\bY_{b_k}=A_{k,b_k}\bg_k+\bW_{b_k}.
	\end{align}
	The estimate of $A_{k,b_k}$ is obtained as
	\begin{align}
	\hat{A}_{k,b_k}=\bg_k^{\dagger}\bY_{b_k}/C.
	\end{align}
	Define the channel estimation error as
	\begin{align}\
	A_k=\hat{A}_{k,b_k}-A_{k,b_k},
	\end{align}
	where $A_{k,b_k}$ is given by \eqref{eq:Akb}.
	
	In iteration $t\geq 2$, with successive cancellation, the updated frequency value at subcarrier $b$ is
	\begin{align}
	\bY_{b} 
	&=\sum_{k \in \mathcal{K} \backslash \mathcal{S}(t-1): b \in \mathcal{B}_k} A_{k,b} \bg_k +  \bW_{b} \nonumber\\
	&\quad- \sum_{\ell \in \mathcal{S}(t-1): b \in \mathcal{B}_{\ell}}\left(\hat{A}_{\ell,b_{\ell}}-A_{\ell,b_{\ell}}e^{\frac{\iota 2\pi(b-b_{\ell})(\hat{\tau}_{\ell}-\tau_{\ell})}{B\Ts}}\right)\nonumber\\
	&\qquad\cdot e^{-\frac{\iota 2\pi (b-b_{\ell})\hat{\tau}_{\ell}}{B\Ts}}\bg_{\ell},
	\end{align}
	where 
	\begin{align}
	\hat{A}_{\ell,b_{\ell}}=	\hat{a}_{\ell} e^{-\frac{\iota 2 \pi b_{\ell}\hat{\tau}_{\ell}  }{B\Ts}},
	\end{align}
	and is how we obtain $\hat{a}_{\ell}$ in \eqref{eq:channel_estimate}.
	Let $\epsilon_{\ell}=\hat{\tau}_{\ell}-\tau_{\ell}$ denote the delay estimation error. Assume $|\epsilon_{\ell}|\leq \rho(\log K)^{-2}$, which decays to 0 as $K\rightarrow\infty$. When $K$ is large, by Taylor expansion,
	\begin{align}
	e^{\frac{\iota 2\pi(b-b_{\ell})(\hat{\tau}_{\ell}-\tau_{\ell})}{B{\Ts}}}&=1+\frac{\iota 2\pi(b-b_{\ell})(\hat{\tau}_{\ell}-\tau_{\ell})}{B{\Ts}}\nonumber\\
	&\quad+O\left(\frac{(b-b_{\ell})^2(\hat{\tau}_{\ell}-\tau_{\ell})^2}{B^2}\right).
	\end{align}
	Since $|b-b_{\ell}|\leq B$, there exists  $\kappa=O(1)$ so that for $K\geq K_0^{\text{r}_1}$ with a sufficiently large $K_0^{\text{r}_1}$, we can write
	\begin{align}
	e^{\frac{\iota 2\pi(b-b_{\ell})(\hat{\tau}_{\ell}-\tau_{\ell})}{B{\Ts}}}=1+\frac{\iota 2\pi(b-b_{\ell})(\hat{\tau}_{\ell}-\tau_{\ell})}{B{\Ts}}+\frac{\kappa}{(\log K)^4}.
	\end{align}
	We hence have
	\begin{align}
	\bY_b
	&=\sum_{k \in \mathcal{K} \backslash \mathcal{S}(t-1): b \in \mathcal{B}_k} A_{k,b} \bg_k +  \bW_{b} \nonumber\\
	&\quad- \sum_{\ell \in \mathcal{S}(t-1): b \in \mathcal{B}_{\ell}}A_{\ell}e^{-\frac{\iota 2\pi (b-b_{\ell})\hat{\tau}_{\ell}}{B\Ts}}\bg_{\ell}\nonumber\\
	&\quad+ \sum_{\ell \in \mathcal{S}(t-1): b \in \mathcal{B}_{\ell}}A_{\ell,b_{\ell}}\left(\frac{\iota 2\pi (b-b_{\ell})\epsilon_{\ell}}{B{\Ts}}+\frac{\kappa}{(\log K)^4}\right)\nonumber\\
	&\qquad\cdot e^{-\frac{\iota 2\pi (b-b_{\ell})\hat{\tau}_{\ell}}{B\Ts}}\bg_{\ell}.\label{eq:calG}
	\end{align}
	
	Suppose now some subcarrier $b_k$ is a singleton due to device $k$  in iteration $t$, and is used to recover device $k$'s index. The channel estimation error is calculated as
	\begin{align}
	A_k &= {\bg_k^{\dagger} \bY_{b_k}}/{C} - A_{k,b_k}  \\
	\label{eq:order_term}& = \frac{\bg_k^{\dagger}}{C} \Bigg(A_{k,b_k} \bg_k + \bW_{b_k} - \sum_{\ell \in \mathcal{S}(t-1): b_k\in \mathcal{B}_{\ell}}A_{\ell}e^{-\frac{\iota 2\pi (b_k-b_{\ell})\hat{\tau}_{\ell}}{B\Ts}}\bg_{\ell}\nonumber\\
	&\quad+ \sum_{\ell \in \mathcal{S}(t-1): b_k\in \mathcal{B}_{\ell}}A_{\ell,b_{\ell}}\left(\frac{\iota 2\pi (b_k-b_{\ell})\epsilon_{\ell}}{B{\Ts}}+\frac{\kappa}{(\log K)^4}\right)\nonumber\\
	&\qquad\cdot e^{-\frac{\iota 2\pi (b_k-b_{\ell})\hat{\tau}_{\ell}}{B\Ts}}\bg_{\ell}\Bigg)- A_{k,b_k} \\
	\label{eq:pk_last}& = \sum_{\ell \in \mathcal{S}(t-1): b_k\in \mathcal{B}_{\ell}}{A_{\ell,b_{\ell}}}\left(\frac{\iota 2\pi (b_k-b_{\ell})\epsilon_{\ell}}{B{\Ts}}+\frac{\kappa}{(\log K)^4}\right)\nonumber\\
	&\qquad\cdot e^{-\frac{\iota 2\pi (b_k-b_{\ell})\hat{\tau}_{\ell}}{B\Ts}}\bg_k^{\dagger} \bg_{\ell}/C+{{\bg}_k^{\dagger} {\bW}_{b_k}}/{C} \nonumber\\
	&\quad-  \sum_{\ell \in \mathcal{S}(t-1): b_k \in \mathcal{B}_{\ell}}  {A_{\ell}} e^{-\frac{\iota 2\pi (b_k-b_{\ell})\hat{\tau}_{\ell}}{B\Ts}}\bg_k^{\dagger} \bg_{\ell}/C,
	\end{align}
	where we use the fact that 
	$\bg_k^{\dagger}\bg_k=C$ to obtain \eqref{eq:pk_last}.
	
	Using recursion, the estimation error of $A_{k,b}$ for $k \in \mathcal{S}(t) \backslash \mathcal{S}(t-1)$ is calculated as
	\begin{align}
	A_k	&=\sum_{\ell_0\in\mathcal{P}(k),\ell_0\in \mathcal{S}(t-1)}{A_{\ell_0,b_{\ell_0}}}\left(\frac{\iota 2\pi (b-b_{\ell_0})\epsilon_{\ell_0}}{B{\Ts}}+\frac{\kappa}{(\log K)^4}\right)\nonumber\\
	&\qquad\cdot e^{-\frac{\iota 2\pi (b-b_{\ell_0})\hat{\tau}_{\ell_0}}{B\Ts}}\bg_k^{\dagger} \bg_{\ell_0}/{C}\nonumber\\
	&\quad+\sum_{(\ell_0,\cdots,\ell_i)\in \mathcal{P}(k),\ell_0\in \mathcal{S}(t-1)}e_{\ell_0}\left(-{\bg^{\dagger}_k\bg_{\ell_{i}}}/{C}\right)\nonumber\\
	&\qquad\cdot\left(-{\bg^{\dagger}_{\ell_{i}}\bg_{\ell_{i-1}}}/{C}\right)\cdots \left(-{\bg^{\dagger}_{\ell_1}\bg_{\ell_0}}/{C}\right)\nonumber\\
	&\qquad\cdot e^{-\frac{\iota 2\pi\left((b-b_{\ell_i})\hat{\tau}_{\ell_i}+(b_{\ell_i}-b_{\ell_{i-1}})\hat{\tau}_{\ell_{i-1}}+\cdots+(b_{\ell_1}-b_{\ell_0})\hat{\tau}_{\ell_0}\right)}{B\Ts}}\nonumber\\
	&\quad+\sum_{(\ell_0,\cdots,\ell_i)\in\mathcal{P}(k),\ell_0\in\mathcal{S}(t-1),i\geq 1}A_{\ell_0,b_{\ell_0}}\nonumber\\
	&\qquad\cdot\left(\frac{\iota 2\pi (b_{\ell_1}-b_{\ell_0})\epsilon_{\ell_0}}{B{\Ts}}+\frac{\kappa}{(\log K)^4}\right)\left({\bg^{\dagger}_k\bg_{\ell_{i}}}/{C}\right)\nonumber\\
	&\qquad\cdot\left(-{\bg^{\dagger}_{\ell_{i}}\bg_{\ell_{i-1}}}/{C}\right)\cdots\left(-{\bg^{\dagger}_{\ell_1}\bg_{\ell_0}}/{C}\right)\nonumber\\
	& \qquad\cdot e^{-\frac{\iota 2\pi\left((b-b_{\ell_i})\hat{\tau}_{\ell_i}+(b_{\ell_i}-b_{\ell_{i-1}})\hat{\tau}_{\ell_{i-1}}+\cdots+(b_{\ell_1}-b_{\ell_0})\hat{\tau}_{\ell_0}\right)}{B\Ts}}\nonumber\\
	&\quad+e_k,
	\end{align}
	where $\mathcal{P}(k)=\{(\ell_0,\cdots,\ell_i):\ell_0,\cdots,\ell_i,k\text{ is a path of}\\\text{devices in the error propagation graph}\}$, and  {$e_{\ell_0}$ is given by \eqref{eq:e_k}}. When $(\ell_0)\in\mathcal{P}(k),\ell_0\in\mathcal{S}(t-1)$, $(-\bg^{\dagger}_k\bg_{\ell_{i}}/C)(-\bg^{\dagger}_{\ell_{i}}\bg_{\ell_{i-1}}/C)\cdots(-\bg^{\dagger}_{\ell_1}\bg_{\ell_0}/C)$ reduces to $-\bg^{\dagger}_k\bg_{\ell_0}/C$. Therefore, the frequency value at subcarrier $b$ in iteration $t$ can be expressed by \eqref{eq:Yb},
	where 
	\begin{align}
	\bV_b&=\sum_{\ell_0\in\mathcal{P}'(b),\ell_0\in \mathcal{S}(t-1)}A_{\ell_0,b_{\ell_0}}\left(\frac{\iota 2\pi (b-b_{\ell_0})\epsilon_{\ell_0}}{B{\Ts}}+\frac{\kappa}{(\log K)^4}\right)\nonumber\\
	&\qquad\cdot e^{-\frac{\iota 2\pi (b-b_{\ell_0})\hat{\tau}_{\ell_0}}{B\Ts}}\bg_{\ell_0}\nonumber\\
	&\quad-\sum_{(\ell_0,\cdots,\ell_i)\in \mathcal{P}'(b),\ell_0\in \mathcal{S}(t-1)}e_{\ell_0}\nonumber\\
	&\qquad\cdot\left(-{\bg^{\dagger}_{\ell_{i}}\bg_{\ell_{i-1}}}/{C}\right)\cdots\left(-{\bg^{\dagger}_{\ell_1}\bg_{\ell_0}}/{C}\right)\bg_{\ell_i}\nonumber\\
	& \qquad\cdot e^{-\frac{\iota 2\pi\left((b-b_{\ell_i})\hat{\tau}_{\ell_i}+(b_{\ell_i}-b_{\ell_{i-1}})\hat{\tau}_{\ell_{i-1}}+\cdots+(b_{\ell_1}-b_{\ell_0})\hat{\tau}_{\ell_0}\right)}{B\Ts}}\nonumber\\
	&\quad+\sum_{(\ell_0,\cdots,\ell_i)\in\mathcal{P}'(b),\ell_0\in\mathcal{S}(t-1),i\geq 1}A_{\ell_0,b_{\ell_0}}\nonumber\\
	&\qquad\cdot\left(\frac{\iota 2\pi (b_{\ell_1}-b_{\ell_0})\epsilon_{\ell_0}}{B{\Ts}}+\frac{\kappa}{(\log K)^4}\right)\nonumber\\
	&\qquad\cdot\left(-{\bg^{\dagger}_{\ell_{i}}\bg_{\ell_{i-1}}}/{C}\right)\cdots\left(-{\bg^{\dagger}_{\ell_1}\bg_{\ell_0}}/{C}\right)\bg_{\ell_i}\nonumber\\
	& \qquad\cdot e^{-\frac{\iota 2\pi\left((b-b_{\ell_i})\hat{\tau}_{\ell_i}+(b_{\ell_i}-b_{\ell_{i-1}})\hat{\tau}_{\ell_{i-1}}+\cdots+(b_{\ell_1}-b_{\ell_0})\hat{\tau}_{\ell_0}\right)}{B\Ts}},
	\end{align}
	and $\mathcal{P}'(b)=\{(\ell_0,\cdots,\ell_i):\ell_0,\cdots,\ell_i\text{ is a path of devices}\\\text{leading to subcarrier }b\text{ in the error propagation graph}\}$. Note that when $(\ell_0)\in\mathcal{P}'(b),\ell_0\in\mathcal{S}(t-1)$, $(-\bg^{\dagger}_{\ell_{i}}\bg_{\ell_{i-1}}/C)\cdots(-\bg^{\dagger}_{\ell_1}\bg_{\ell_0}/C)\bg_{\ell_i}$ reduces to $\bg_{\ell_0}$.
	
	Suppose $G\in\mathcal{G}$, then $|\mathcal{P}'(b)|\leq 2$. Moreover, by Proposition \ref{prop:prop1}, the number of left nodes in each component is less than $\alpha D\log(\beta_0K)/(D-1)$, which indicates that $|\mathcal{S}(t-1)|\leq\alpha D\log(\beta_0 K)/(D-1)\leq \alpha D\log K/(D-1)+2\alpha\log\beta_0$. We then upper bound the coefficients of $\epsilon_{\ell_0}$ and $e_{\ell_0}$ in the three summation terms of $\bV_b$.
	
	For each entry of $\bV_b$, since $A_{\ell_0,b_{\ell_0}}=a_{\ell_0}e^{-\frac{\iota 2\pi b_{\ell_0}\tau_{\ell_0}}{B\Ts}}$ where $\ua\leq |a_{\ell_0}|\leq\bar{a}$, the delay estimation error $\epsilon_{\ell_0}$ in the first summation term has the coefficient satisfying 
	\begin{align}
	&\left\vert A_{\ell_0,b_{\ell_0}}\left(\frac{\iota 2\pi (b-b_{\ell_0})\epsilon_{\ell_0}}{B{\Ts}}+\frac{\kappa}{(\log K)^4}\right)e^{-\frac{\iota 2\pi (b-b_{\ell_0})\hat{\tau}_{\ell_0}}{B\Ts}}\right\vert\nonumber\\
	&\quad\leq\frac{2\pi\bar{a}|\epsilon_{\ell_0}|}{{\Ts}}+\frac{\bar{a}\kappa}{(\log K)^4}.
	\end{align}
	Since $|\exp(\iota x)|=1,\forall x\in\mathbb{R}$, we neglect the exponent terms when bounding the amplitude in the following for brevity. Since the entries of the design parameter $\bg_{\ell}$ are i.i.d. BPSK symbols and $\bg_{\ell}\in\mathbb{R}^C$, $|-\bg^{\dagger}_{\ell_i}\bg_{\ell_{i-1}}/C|\leq 1$. The delay estimation error in the third summation term has the coefficient satisfying
	\begin{align}
	&\left\vert A_{\ell_0,b_{\ell_0}}\left(\frac{\iota 2\pi (b_{\ell_1}-b_{\ell_0})\epsilon_{\ell_0}}{B{\Ts}}+\frac{\kappa}{(\log K)^4}\right)\right.\nonumber\\
	&\quad\left.\cdot\left(-{\bg^{\dagger}_{\ell_{i}}\bg_{\ell_{i-1}}}/{C}\right)\cdots\left(-{\bg^{\dagger}_{\ell_1}\bg_{\ell_0}}/{C}\right)\right\vert\leq\frac{2\pi\bar{a}|\epsilon_{\ell_0}|}{{\Ts}}+\frac{\bar{a}\kappa}{(\log K)^4}.
	\end{align}
	Moreover, $e_{\ell_0}$ has the coefficient satisfying 
	\begin{align}
	&\left\vert\left(-{\bg^{\dagger}_{\ell_{i}}\bg_{\ell_{i-1}}}/{C}\right)\cdots\left(-{\bg^{\dagger}_{\ell_1}\bg_{\ell_0}}/{C}\right)\right\vert\leq 1.
	\end{align}
	Combining with the assumption that $e_{\ell_0}$ is upper bounded by \eqref{eq:ub_ek} and $\epsilon_{\ell_0}$ is upper bounded by \eqref{eq:ub_epsilonk}, each entry of $\bV_b$ is upper bounded by 
	\begin{align}
	|V_b^c| &\leq |\mathcal{P}'(b)| |\mathcal{S}(t-1)|\left(\frac{4\pi\bar{a}|\epsilon_{\ell_0}|}{{\Ts}}+|e_{\ell_0}|+\frac{2\bar{a}\kappa}{(\log K)^4}\right)\\
	&\leq \frac{2\alpha D\log(\beta_0 K)}{D -1}\left(\frac{4\pi\bar{a}\rho/{\Ts}+\varrho}{(\log K)^2}+\frac{2\bar{a}\kappa}{(\log K)^4}\right).
	\end{align}
	Hence, for $K\geq K_0^{\text{r}_2}$    with a sufficiently large $K_0^{\text{r}_2}$, 
	\begin{align}
	|V_b^c| &\leq \frac{4\alpha D\log(\beta_0 K)}{D -1}\frac{4\pi\bar{a}\rho/{\Ts}+\varrho}{(\log K)^2}.\label{eq:ub_vbc_bound}
	\end{align}
	
	Plugging \eqref{eq:ub_varrho} and \eqref{eq:ub_rho} into \eqref{eq:ub_vbc_bound} and using the fact $\log(\beta_0 K)\leq(1+\log\beta_0)\log K$ yield \eqref{eq:ub_vbc}. Let $K_0^{\text{r}}=\max\{K_0^{\text{r}_1},K_0^{\text{r}_2}\}$. Hence the proof of Proposition \ref{prop:prop3}.

	\subsection{Proof of Proposition~\ref{prop:prop4}}\label{app:prop4}
	The robust subcarrier detection and the error due to noise have been analyzed in \cite{chen2022asynchronous}. Moreover, $\eta$ can be picked as a constant that satisfies
    \begin{align}
    \eta\geq\frac{32\sigma^2\lceil\beta_1\log K\rceil\log_e(K)}{\beta_0K},
    \end{align}
    whose right hand side vanishes as $K\rightarrow\infty$. We hence can choose $\eta=\ua^2$. Note that in \cite{chen2022asynchronous}, $\lceil \log N\rceil$ OFDM symbols are used for phase estimation, which can be reduced to 1 without affecting the desired estimation accuracy. 
	
	We now bound the delay estimation error. The last subframe lasts for time duration $C_2{\Ts }$, where $C_2$ is given by~\eqref{eq:para_C2}
	with $\beta_2\geq 98304{{\Ts }^2}/{\rho^2}$. As a result, $I=C_2-M$ is
		\begin{align}\label{eq:I}
			I =\lceil \beta_2(\bar{a}/\ua)^2(\log K)^4K\log(KM+1)\rceil.
		\end{align}

		We use a two-step process for estimating delays. First, we use the crude estimation to find a slot of length ${\Ts}$ that likely contains the delay $\tau_k$ with a high probability. If the crude estimation is successful, we then use the fine estimation to find a slot of length no greater than $\frac{\rho}{2(\log K)^2}$ that contains $\tau_k$ with a high probability. If both steps are successful, we can pick any $\hat{\tau}_k$ from the slot given in \eqref{eq:t_fine} to have that $|\hat{\tau}_k-\tau_k|\leq\frac{\rho}{(\log K)^2}$.
		
		Device $k$'s delay can be an arbitrary real number in $[0,M{\Ts }]$. Without loss of generality, we assume $\tau_k=\psi/4$ with $\psi$ given in \eqref{eq:fine_length}. Note that when $\tau_k=0$ or $\tau_k=M{\Ts }$, the delay can only be overestimated or underestimated, respectively. However, this will not change the conclusion or the proof.

		\subsubsection{Crude estimation}
		In the first step, we use a threshold-based testing to find a slot of $\tau$ having length ${\Ts }$ such that $|\tau-\tau_k|\leq {\Ts }$ with probability $1-O(K^{-2})$.
		
		The main fact utilized is that if an unknown $\tau_k\in[i{\Ts },(i+1){\Ts })$ with some $i \in\{ 0,1,\cdots, M-1\}$, then
		\begin{align}
		&|i'{\Ts }-\tau_k|\leq {\Ts }/2,  i'=i,\text{or } i' = i+1, \text{or } i' = i,i+1;\\
		&|i'{\Ts }-\tau_k|\geq {\Ts }, i'\neq i,i+1.
		\end{align}
        Additionally, when $|\tau-\tau_k|\leq T/2$, it is highly likely to have $|{\calT}_k(\tau)|\geq \bar{\calT}$, and when $|\tau-\tau_k|\geq T$, it is highly likely to have $|{\calT}_k(\tau)|< \bar{\calT}$. We thus can locate $\tau_k$ through \eqref{eq:t_crude}. To be more specific, with the assumption $\tau_k=\psi/4\in[0,{\Ts })$, we show that
		\begin{align}
		&\mathsf{P}(|\calT_k(0)|\geq\bar{\calT})\geq1-\frac{16}{K^2},\label{eq:Tk0}\\
		&\mathsf{P}(|\calT_k(\tau)|<\bar{\calT})\geq 1-\frac{12}{MK^2},\quad\forall\tau\in\calH, \tau\neq 0,{\Ts }.\label{eq:Tkt}
		\end{align}
		Denote by $P_{e_1}$ the probability of failure in detecting the desired slot in the crude estimation. It follows that 
		\begin{align}
		P_{e_1}&\leq \sum_{\tau\in\calH,\tau\neq 0, {\Ts }}\Pr\{|\calT_k(\tau)|\geq\bar{\calT}\} + \Pr\{|\calT_k(0)|<\bar{\calT}\} \\
		&\leq\frac{28}{K^2}.
		\end{align}
  
		In the following, we elaborate the proof for \eqref{eq:Tk0} and \eqref{eq:Tkt}. When $\tau\in\calH,\tau\neq 0,{\Ts }$, we can write
		\begin{align}
		\mathsf{P}\left\{|\calT_k(\tau)|\geq\bar{\calT}\right\}
		&\leq\Pr\left\{\vert\Re\{\calT_k(\tau)\}\vert\geq\bar{\calT}/2\right\}\nonumber\\
		&\quad+\Pr\left\{\vert\Im\{\calT_k(\tau)\}\vert\geq\bar{\calT}/2\right\}.\label{eq:Ik_T}
		\end{align}
		To proceed, we split $\vert\Re\{\calT_k(\tau)\}\vert$ and have
		\begin{align}
		&\Pr\left\{\vert\Re\{\calT_k(\tau)\}\vert\geq\bar{\calT}/2\right\}\nonumber\\
		&\leq \mathsf{P}\left\{\left\vert \sum_{k\in\mathcal{K}}\sum_{i=0}^{I-1}\Re\{a_k\}\Delta_k R_{i,k}\right\vert\geq \bar{\calT}/8\right\}\nonumber\\
		&\quad+\mathsf{P}\left\{\left\vert \sum_{k\in\mathcal{K}}\sum_{i=0}^{I-1}\Re\{a_k\}({\Ts }-\Delta_k){R'}_{i,k}\right\vert\geq \bar{\calT}/8\right\}\nonumber\\
		&\quad+\Pr\left\{\left\vert\sum_{i=0}^{I-1}\Re\{Z_i\}\right\vert\geq \bar{\calT}/4\right\}.\label{eq:Tk_T_re}
		\end{align}
		For the first term on the right hand side of \eqref{eq:Tk_T_re}, recall that $R_{i,k}$ is a BPSK symbol, and hence is a $1$-subGaussian with zero mean. By properties of subGaussian random variables \cite{chen2022asynchronous,rivasplata2012subgaussian}, $\sum_{k\in\calK}\sum_{i=0}^{I-1}\Re\{a_k\}\Delta_kR_{i,k}$ is $\sqrt{I\sum_{k\in\calK}(\Re\{a_k\}\Delta_k)^2}$-subGaussian variable with zero mean. Moreover, we have 
		\begin{align}
		&\Pr\left\{\left\vert \sum_{k\in\mathcal{K}}\sum_{i=0}^{I-1}\Re\{a_k\}\Delta_k R_{i,k}\right\vert\geq \bar{\calT}/8\right\}\nonumber\\
		&\leq 2\exp\left(-\frac{\bar{\calT}^2}{128 I\sum_{k\in\calK}(\Re\{a_k\}\Delta_k)^2}\right)\label{eq:sub_tail}\\
		&\leq 2\exp\left(-\frac{\ua^2 I}{2048 K\bar{a}^2}\right)\label{eq:adel}\\
		&\leq 2\exp\left(-\frac{48{\Ts }^2(\log K)^4\log(KM+1)}{\rho^2}\right)\label{eq:Ibeta}\\
		&\leq 2\exp(-2\log(MK))\label{eq:loose}\\
		&\leq \frac{2}{MK^2},\label{eq:Tk_T_1}
		\end{align}
		where \eqref{eq:sub_tail} is obtained according to the tail probability bound of subGaussian random variables, \eqref{eq:adel} is because that $|\Re\{a_k\}|\leq|a_k|\leq\bar{a}$ and $\Delta_k<{\Ts }$, and \eqref{eq:Ibeta} is due to \eqref{eq:I} with $\beta_2\geq 98304{{\Ts }^2}/{\rho^2}$. Similarly, we also have
		\begin{align}
		\Pr\left\{\left\vert \sum_{k\in\mathcal{K}}\sum_{i=0}^{I-1}\Re\{a_k\}({\Ts }-\Delta_k) {R'}_{i,k}\right\vert\geq \bar{\calT}/8\right\}\leq \frac{2}{MK^2}.\label{eq:Tk_T_2}
		\end{align}
		For the last term on the right hand side of \eqref{eq:Tk_T_re}, recall that $\Re\{Z_i\}$ is i.i.d. Gaussian variables with zero mean and variance ${\Ts }\sigma^2$. Therefore,
		\begin{align}
		\Pr\left\{\left\vert\sum_{i=0}^{I-1}\Re\{Z_i\}\right\vert\geq \bar{\calT}/4\right\}&=2Q\left(\frac{\bar{\calT}}{4\sqrt{I{\Ts }\sigma^2}}\right)\label{eq:Qfunc}\\
		&\leq 2\exp\left(-\frac{\bar{\calT}^2}{32 I{\Ts }\sigma^2}\right)\label{eq:Q_ieq}\\
		&=2\exp\left(-\frac{\ua^2{\Ts } I}{512\sigma^2}\right)\\
		&\leq \frac{2}{MK^2},\label{eq:Tk_T_3}
		\end{align}
		where the $Q$ function in \eqref{eq:Qfunc} is the tail distribution function of the standard normal distribution, \eqref{eq:Q_ieq} is due to $Q(x)\leq\exp(-x^2/2)$, and \eqref{eq:Tk_T_3} follows from \eqref{eq:I}. Thus, by \eqref{eq:Tk_T_re},\eqref{eq:Tk_T_1},\eqref{eq:Tk_T_2}, and \eqref{eq:Tk_T_3}, we have
		\begin{align}
		\Pr\left\{\vert\Re\{\calT_k(\tau)\}\vert\geq\bar{\calT}/2\right\}\leq\frac{6}{MK^2}.\label{eq:Tk_T_Re_up}
		\end{align}
		Following the similar derivations, we can obtain $\Pr\left\{\vert\Im\{\calT_k(\tau)\}\vert\geq\bar{\calT}/2\right\}\leq{6}/{(MK^2)}$. It follows that
		\begin{align}
		\Pr\{|\calT_k(\tau)|\geq\bar{\calT}\}\leq\frac{12}{MK^2},\quad\tau\in\calH,\tau\neq 0,{\Ts }.
		\end{align}
		
		When $\tau=0$, we have $|\tau-\tau_k|\leq{\Ts }/2$. We thus have
		\begin{align}
		&\mathsf{P}\left\{|\calT_k(0)|\leq\bar{\calT}\right\}\nonumber\\
		&=\mathsf{P}\left\{\left\vert 	
		a_k \left(({\Ts }-\tau_k)I+\tau_k\sum_{i=0}^{I-1}R_{i,k}\right)+\sum_{i=0}^{I-1}Z_i\right.\right.\nonumber\\
		&\left.\left.\qquad+\sum_{p\in\mathcal{K}\backslash k}a_p\left(\Delta_p\sum_{i=0}^{I-1}R_{i,p}+({\Ts }-\Delta_p)\sum_{i=0}^{I-1}{R'}_{i,p}\right)
		\right\vert\leq \bar{\calT}\right\}\\
		&\leq \mathsf{P}\left\{\left\vert a_k ({\Ts }-\tau_k)I\right\vert-\left\vert a_k\tau_k\sum_{i=0}^{I-1}R_{i,k} +\sum_{i=0}^{I-1}Z_i\right.\right.\nonumber\\
		&\qquad\left.\left.+\sum_{p\in\mathcal{K}\backslash k}a_p\left(\Delta_p\sum_{i=0}^{I-1}R_{i,p}+({\Ts }-\Delta_p)\sum_{i=0}^{I-1}{R'}_{i,p}\right)\right\vert\leq \bar{\calT}\right\}\\
		&\leq\mathsf{P}\left\{\left\vert a_k\tau_k\sum_{i=0}^{I-1}R_{i,k}+\sum_{i=0}^{I-1}Z_i+\sum_{p\in\mathcal{K}\backslash k}a_p\Delta_p\sum_{i=0}^{I-1}R_{i,p}\right.\right.\nonumber\\
		&\qquad\left.\left.+\sum_{p\in\mathcal{K}\backslash k}a_p({\Ts }-\Delta_p)\sum_{i=0}^{I-1}{R'}_{i,p}\right\vert\geq \bar{\calT}\right\}\label{eq:amp_ass}\\
		&\leq\mathsf{P}\left\{\left\vert \Re\left\{a_k\tau_k\sum_{i=0}^{I-1}R_{i,k}+\sum_{i=0}^{I-1}Z_i+\sum_{p\in\mathcal{K}\backslash k}a_p\Delta_p\sum_{i=0}^{I-1}R_{i,p}\right.\right.\right.\nonumber\\
		&\qquad\left.\left.\left. +\sum_{p\in\mathcal{K}\backslash k}a_p({\Ts }-\Delta_p)\sum_{i=0}^{I-1}{R'}_{i,p}\right\}\right\vert\geq \frac{\bar{\calT}}{2}\right\}\nonumber\\
		&\quad+\mathsf{P}\left\{\left\vert \Im\left\{a_k\tau_k\sum_{i=0}^{I-1}R_{i,k}+\sum_{i=0}^{I-1}Z_i+\sum_{p\in\mathcal{K}\backslash k}a_p\Delta_p\sum_{i=0}^{I-1}R_{i,p}\right.\right.\right.\nonumber\\
		&\qquad\left.\left.\left. +\sum_{p\in\mathcal{K}\backslash k}a_p({\Ts }-\Delta_p)\sum_{i=0}^{I-1}{R'}_{i,p}\right\}\right\vert\geq \frac{\bar{\calT}}{2}\right\},
		\end{align}
		where \eqref{eq:amp_ass} is because of the assumption $|a_k|\geq a$, and $|\tau_k|\leq{\Ts }/2$.
		Following the similar derivations for \eqref{eq:Tk_T_re}, we have 
		\begin{align}
		&\mathsf{P}\left\{\left\vert \Re\left\{a_k\tau_k\sum_{i=0}^{I-1}R_{i,k}+\sum_{i=0}^{I-1}Z_i+\sum_{p\in\mathcal{K}\backslash k}a_p\Delta_p\sum_{i=0}^{I-1}R_{i,p}\right.\right.\right.\nonumber\\
		&\qquad\left.\left.\left. +\sum_{p\in\mathcal{K}\backslash k}a_p({\Ts }-\Delta_p)\sum_{i=0}^{I-1}{R'}_{i,p}\right\}\right\vert\geq \frac{\bar{\calT}}{2}\right\}\nonumber\\
		&\leq \mathsf{P}\left\{\left\vert \Re\left\{a_k\right\}\tau_k\sum_{i=0}^{I-1}R_{i,k}\right\vert\geq \frac{\bar{\calT}}{8}\right\}+\mathsf{P}\left\{\left\vert \sum_{i=0}^{I-1}\Re\left\{Z_i\right\}\right\vert\geq \frac{\bar{\calT}}{8}\right\}\nonumber\\
		&\quad+\mathsf{P}\left\{\left\vert \sum_{p\in\mathcal{K}\backslash k}\Re\left\{a_p\right\}\Delta_p\sum_{i=0}^{I-1}R_{i,p}\right\vert\geq \frac{\bar{\calT}}{8}\right\}\nonumber\\
		&\quad+\mathsf{P}\left\{\left\vert\sum_{p\in\mathcal{K}\backslash k}\Re\left\{a_p\right\}({\Ts }-\Delta_p)\sum_{i=0}^{I-1}{R'}_{i,p}\right\vert\geq \frac{\bar{\calT}}{8}\right\}\\
		&\leq \frac{8}{K^2}.
		\end{align}
		It can also be verified that the magnitude of the imaginary part has the same probability bound. We thus have
		\begin{align}
		\Pr\{|\calT_k(0)|\leq\bar{\calT}\}\leq \frac{16}{K^2}.
		\end{align} 
		
		\subsubsection{Refined estimation.}
		Suppose we succeed in finding a slot of length ${\Ts }$ that contains $\tau_k$ in the crude estimation step, and hence have an offset $\lambda$. Now we find a slot of $\tau$ having length $\psi$ such that $|\tau-\tau_k|\leq\rho(\log K)^{-2}$  with probability $1-O(K^{-2})$.
		
		The main fact utilized is that if an unknown $\tau_k\in[i\psi+\lambda,(i+1)\psi+\lambda\})$ with some $ i\in\left\{0,1,\cdots,\left\lceil \frac{2{\Ts }(\log K)^2}{\rho}\right\rceil-1\right\}$, then
		\begin{align}
		&\left|i'\psi-\tau_k+\lambda\right|\leq\frac{\psi}{2}\leq \frac{\rho}{4 (\log K)^2},\nonumber\\
		&\qquad\qquad\qquad i'=i,\text{ or }i' = i+1, \text{ or }i'=i,i+1,\\
		&\left|i'\psi-\tau_k+\lambda\right|\geq \psi,\quad i'\neq i,i+1.
		\end{align}
		Then, with a large probability, either $\tau=i\psi+\lambda$ or $\tau = (i+1)\psi+\lambda$ has the largest statistic values among $\tau+\lambda,\forall\tau\in\calH'$. According to \eqref{eq:t_fine}, whichever $|\calT_k(i\psi+\lambda)|$ or $|\calT_k((i+1)\psi+\lambda)|$ is the largest one,  $[i\psi+\lambda,(i+1)\psi+\lambda)$ is contained in the found slot, and so is $\tau_k$.
		
		Since $\tau_k=\psi/4$, given the crude estimation is successful, we have $\lambda=0$. Then we show that
		\begin{align}\label{eq:Tkcompare}
		\Pr\left\{\left|\calT_k\left(0\right)\right|\leq \left|\calT_k\left(i'\psi\right)\right|\right\}\leq\frac{32}{MK^3},\forall i'\psi\in\calH',i'\geq 2.
		\end{align}
		Denote by $P_{e_2}$ the probability of failure in detecting the desired slot in fine estimation given that crude estimation successfully finds the slot of length ${\Ts }$ containing $\tau_k$. We have
		\begin{align}
		P_{e_2}&\leq\sum_{i'\psi\in\calH',i'\geq 2}\Pr\left\{\left|\calT_k\left(0\right)\right|\leq \left|\calT_k\left(i'\psi\right)\right|\right\}\\
		&\leq \frac{32}{MK^3}\left\lceil\frac{2{\Ts }(\log K)^2}{\rho}\right\rceil\\
		&\leq \frac{128{\Ts }}{\rho MK^2}.
		\end{align}
		
		In the following, we elaborate the proof for \eqref{eq:Tkcompare}. Notice that for all $i'\geq 2$ and $i'\psi\in\calH'$, we have $\left|i'\psi-\tau_k\right|\geq \psi$. With $i'=0$, we have $\tau_k\leq \frac{\psi}{2}$. Thus, for $i'\geq 2$,
		\begin{align}
		&\Pr\left\{\left|\calT_k\left(0\right)\right|\leq \left|\calT_k\left(i'\psi\right)\right|\right\}\nonumber\\
		&=\Pr\left\{\left|a_k\left(({\Ts }-\tau_k) I+\tau_k\sum_{i=0}^{I-1}R_{i,k}\right)+\sum_{i=0}^{I-1}Z_{i_1}\right.\right.\nonumber\\
		&\qquad\left.\left.+\sum_{p\in\mathcal{K}\backslash k}a_p\left(\Delta_{p_1}\sum_{i=0}^{I-1}R_{i,{p_1}}+({\Ts }-\Delta_{p_1})\sum_{i=0}^{I-1}{R'}_{i,{p_1}}\right)\right|\right.\nonumber\\
		&\qquad\left.\leq\left|a_k ({\Ts }-|i'\psi-\tau_k|)I+a_k |i'\psi-\tau_k|\sum_{i=0}^{I-1}R_{i,k}\right.\right.\nonumber\\
		&\left.\left.\qquad+\sum_{i=0}^{I-1}Z_{i_2}+\sum_{p\in\mathcal{K}\backslash k}a_p\Delta_{p_2}\sum_{i=0}^{I-1}R_{i,{p_2}}\right.\right.\nonumber\\
        &\left.\left.\qquad+\sum_{p\in\mathcal{K}\backslash k}a_p({\Ts }-\Delta_{p_2})\sum_{i=0}^{I-1}{R'}_{i,{p_2}}\right|\right\}\\
		&\leq\Pr\left\{|a_k|({\Ts }-\tau_k) I-\left|a_k\tau_k\sum_{i=0}^{I-1}R_{i,k}+\sum_{i=0}^{I-1}Z_{i_1}\right.\right.\nonumber\\
		&\qquad\left.\left.+\sum_{p\in\mathcal{K}\backslash k}a_p\left(\Delta_{p_1}\sum_{i=0}^{I-1}R_{i,{p_1}}+({\Ts }-\Delta_{p_1})\sum_{i=0}^{I-1}{R'}_{i,{p_1}}\right)\right|\right.\nonumber\\
	   &\qquad\left.\leq |a_k| \left({\Ts }-|i'\psi-\tau_k|\right)I+\left|a_k(i'\psi-\tau_k)\sum_{i=0}^{I-1}R_{i,k}\right.\right.\nonumber\\
		&\qquad\left.\left.+\sum_{i=0}^{I-1}Z_{i_2}+\sum_{p\in\mathcal{K}\backslash k}a_p\Delta_{p_2}\sum_{i=0}^{I-1}R_{i,{p_2}}\right.\right.\nonumber\\
        &\qquad\left.\left.+\sum_{p\in\mathcal{K}\backslash k}a_p({\Ts }-\Delta_{p_2})\sum_{i=0}^{I-1}{R'}_{i,{p_2}}\right|\right\}\\
		&\leq\Pr\left\{\left|\sum_{p\in\mathcal{K}\backslash k}a_p\left(\Delta_{p_1}\sum_{i=0}^{I-1}R_{i,{p_1}}+({\Ts }-\Delta_{p_1})\sum_{i=0}^{I-1}{R'}_{i,{p_1}}\right)\right.\right.\nonumber\\
		&\qquad\left.\left.+a_k\tau_k\sum_{i=0}^{I-1}R_{i,k}+\sum_{i=0}^{I-1}Z_{i_1}\right|\right.\nonumber\\
		&\qquad\left.+\left|\sum_{p\in\mathcal{K}\backslash k}a_p\left(\Delta_{p_2}\sum_{i=0}^{I-1}R_{i,{p_2}}+({\Ts }-\Delta_{p_2})\sum_{i=0}^{I-1}{R'}_{i,{p_2}}\right)\right.\right.\nonumber\\
		&\qquad\left.\left. +a_k(i'\psi-\tau_k)\sum_{i=0}^{I-1}R_{i,k}+\sum_{i=0}^{I-1}Z_{i_2}\right|\geq \frac{\ua\psi I}{2}\right\}\label{eq:a_lower}\\
		&\leq \mathsf{P}\left\{\left|\sum_{p\in\mathcal{K}\backslash k}a_p\left(\Delta_{p_1}\sum_{i=0}^{I-1}R_{i,{p_1}}+({\Ts }-\Delta_{p_1})\sum_{i=0}^{I-1}{R'}_{i,{p_1}}\right)\right.\right.\nonumber\\
		&\qquad\left.\left.+a_k\tau_k\sum_{i=0}^{I-1}R_{i,k}+\sum_{i=0}^{I-1}Z_{i_1}\right|\geq  \frac{\ua\psi I}{4}\right\}\nonumber\\
		&\quad+\Pr\left\{\left|\sum_{p\in\mathcal{K}\backslash k}a_p\left(\Delta_{p_2}\sum_{i=0}^{I-1}R_{i,{p_2}}+({\Ts }-\Delta_{p_2})\sum_{i=0}^{I-1}{R'}_{i,{p_2}}\right)\right.\right.\nonumber\\
		&\qquad\left.\left. +a_k(i'\psi-\tau_k)\sum_{i=0}^{I-1}R_{i,k}+\sum_{i=0}^{I-1}Z_{i_2}\right|\geq  \frac{\ua\psi I}{4}\right\}\label{eq:bound_fine},
		\end{align}
		where \eqref{eq:a_lower} utilizes the assumption that $|a_k|\geq a$, and the fact that $|i'\psi-\tau_k|-|\tau_k|\geq\psi/2, i'\geq 2$.
		We can further bound the first term of \eqref{eq:bound_fine} as below
		\begin{align}
		&\mathsf{P}\left\{\left|\sum_{p\in\mathcal{K}\backslash k}a_p\left(\Delta_{p_1}\sum_{i=0}^{I-1}R_{i,{p_1}}+({\Ts }-\Delta_{p_1})\sum_{i=0}^{I-1}{R'}_{i,{p_1}}\right)\right.\right.\nonumber\\
		&\qquad\left.\left.+a_k\tau_k\sum_{i=0}^{I-1}R_{i,k}+\sum_{i=0}^{I-1}Z_{i_1}\right|\geq  \frac{\ua\psi I}{4}\right\}\nonumber\\
		&\leq \mathsf{P}\left\{\left|\Re\left\{\sum_{p\in\mathcal{K}\backslash k}a_p\Delta_{p_1}\sum_{i=0}^{I-1}R_{i,{p_1}}\right.\right.\right.\nonumber\\
        &\qquad\left.\left.\left.+\sum_{p\in\mathcal{K}\backslash k}a_p({\Ts }-\Delta_{p_1})\sum_{i=0}^{I-1}{R'}_{i,{p_1}}\right.\right.\right.\nonumber\\
		&\qquad\left.\left.\left.+a_k\tau_k\sum_{i=0}^{I-1}R_{i,k}+\sum_{i=0}^{I-1}Z_{i_1}\right\}\right|\geq \frac{\ua\psi I}{8}\right\}\nonumber\\
		&\quad + \mathsf{P}\left\{\left|\Im\left\{\sum_{p\in\mathcal{K}\backslash k}a_p\Delta_{p_1}\sum_{i=0}^{I-1}R_{i,{p_1}}\right.\right.\right.\nonumber\\
        &\qquad\left.\left.\left.+\sum_{p\in\mathcal{K}\backslash k}a_p({\Ts }-\Delta_{p_1})\sum_{i=0}^{I-1}{R'}_{i,{p_1}}\right.\right.\right.\nonumber\\
		&\qquad\left.\left.\left.+a_k\tau_k\sum_{i=0}^{I-1}R_{i,k}+\sum_{i=0}^{I-1}Z_{i_1}\right\}\right|\geq \frac{\ua\psi I}{8}\right\}.
		\end{align}
		Similar to previous derivations, 
		\begin{align}
		&\mathsf{P}\left\{\left|\Re\left\{\sum_{p\in\mathcal{K}\backslash k}a_p\left(\Delta_{p_1}\sum_{i=0}^{I-1}R_{i,{p_1}}+({\Ts }-\Delta_{p_1})\sum_{i=0}^{I-1}{R'}_{i,{p_1}}\right)\right.\right.\right.\nonumber\\
		&\qquad\left.\left.\left.+a_k\tau_k\sum_{i=0}^{I-1}R_{i,k}+\sum_{i=0}^{I-1}Z_{i_1}\right\}\right|\geq \frac{\ua\psi I}{8}\right\}\nonumber\\
		&\leq\mathsf{P}\left\{\left|\sum_{p\in\mathcal{K}\backslash k}\sum_{i=0}^{I-1}\Re\{a_p\}\Delta_{p_1}R_{i,{p_1}}\right|\geq \frac{\ua\psi I}{32}\right\}\nonumber\\
		&\quad+\mathsf{P}\left\{\left|\sum_{p\in\mathcal{K}\backslash k}\sum_{i=0}^{I-1}\Re\{a_p\}({\Ts }-\Delta_{p_1}){R'}_{i,{p_1}}\right|\geq \frac{\ua\psi I}{32}\right\}\nonumber\\
		&\quad+\mathsf{P}\left\{\left|\sum_{i=0}^{I-1}\Re\{a_k\}\tau_kR_{i,k}\right|\geq \frac{\ua\psi I}{32}\right\}\nonumber\\
		&\quad+\mathsf{P}\left\{\left|\sum_{i=0}^{I-1}\Re\{Z_{i_1}\}\right|\geq \frac{\ua\psi I}{32}\right\}.
		\end{align}
		By properties of subGaussian variable,
		\begin{align}
		&\Pr\left\{\left|\sum_{p\in\mathcal{K}\backslash k}\sum_{i=0}^{I-1}\Re\left\{a_p\right\}\Delta_{p_1}R_{i,{p_1}}\right|\geq \frac{\ua\psi I}{32}\right\}\nonumber\\
		&\leq 2\exp\left(-\frac{\ua^2\psi^2 I^2}{2048 I\sum_{p\in\calK\backslash k}(\Re\{a_p\}\Delta_{p_2})^2}\right)\\
		&\leq 2\exp\left(-\frac{\ua^2\rho^2I}{32768 \bar{a}^2{\Ts }^2K(\log K)^4}\right)\label{eq:I_order}\\
		&\leq 2\exp\left(-3\log(MK+1)\right)\label{eq:strict}\\
		&\leq\frac{2}{MK^3}\label{eq:finer_re_1},
		\end{align}
		where \eqref{eq:I_order} is because $\psi\geq\frac{\rho}{4(\log K)^2}$, and \eqref{eq:strict} is due to \eqref{eq:I} with $\beta_2\geq 98304{{\Ts }^2}/{\rho^2}$. Similarly, we can obtain that 
		\begin{align}
		&\Pr\left\{\left|\sum_{p\in\mathcal{K}\backslash k}\sum_{i=0}^{I-1}\Re\left\{a_p\right\}({\Ts }-\Delta_{p_1}){R'}_{i,{p_1}}\right|\geq \frac{a\psi I}{32}\right\}\leq\frac{2}{MK^3},\label{eq:finer_re_2}\\
		&\Pr\left\{\left|\sum_{i=0}^{I-1}\Re\left\{a_k\right\} \tau_k R_{i,k}\right|\geq \frac{a\psi I}{32}\right\}\leq\frac{2}{MK^3},\label{eq:finer_re_3}\\
		&\Pr\left\{\left\vert\sum_{i=0}^{I-1}\Re\{Z_{i_1}\}\right\vert\geq \frac{a\psi I}{32}\right\}\leq \frac{2}{MK^3}.\label{eq:finer_re_4}
		\end{align}
		
		%
		
		Combining \eqref{eq:finer_re_1},\eqref{eq:finer_re_2},\eqref{eq:finer_re_3} and \eqref{eq:finer_re_4}, we have
		\begin{align}
		&\mathsf{P}\left\{\left|\Re\left\{\sum_{p\in\mathcal{K}\backslash k}a_p\left(\Delta_{p_1}\sum_{i=0}^{I-1}R_{i,{p_1}}+({\Ts }-\Delta_{p_1})\sum_{i=0}^{I-1}{R'}_{i,{p_1}}\right)\right.\right.\right.\nonumber\\
		&\qquad\left.\left.\left.+a_k\tau_k\sum_{i=0}^{I-1}R_{i,k}+\sum_{i=0}^{I-1}Z_{i_1}\right\}\right|\geq \frac{a\psi I}{8}\right\}\leq\frac{8}{MK^3}.
		\end{align}
		The bound also holds for the imaginary part. Thus,
		\begin{align}
		&\mathsf{P}\left\{\left|\sum_{p\in\mathcal{K}\backslash k}a_p\left(\Delta_{p_1}\sum_{i=0}^{I-1}R_{i,{p_1}}+({\Ts }-\Delta_{p_1})\sum_{i=0}^{I-1}{R'}_{i,{p_1}}\right)\right.\right.\nonumber\\
		&\qquad\left.\left.+a_k\tau_k\sum_{i=0}^{I-1}R_{i,k}+\sum_{i=0}^{I-1}Z_{i_1}\right|\geq  \frac{a\psi I}{4}\right\}\leq\frac{16}{MK^3}
		\end{align}
		
		The second term of \eqref{eq:bound_fine} can also be bounded as 
		\begin{align}
		&\Pr\left\{\left|\sum_{p\in\mathcal{K}\backslash k}a_p\left(\Delta_{p_2}\sum_{i=0}^{I-1}R_{i,{p_2}}+({\Ts }-\Delta_{p_2})\sum_{i=0}^{I-1}{R'}_{i,{p_2}}\right)\right.\right.\nonumber\\
		&\qquad\left.\left. +a_k(i'\psi-\tau_k)\sum_{i=0}^{I-1}R_{i,k}+\sum_{i=0}^{I-1}Z_{i_2}\right|\geq  \frac{a\psi I}{4}\right\}\leq\frac{16}{MK^3}.
		\end{align}

		Hence, we have
		\begin{align}
		\Pr\left\{\left|\calT_k\left(0\right)\right|\leq \left|\calT_k\left(i'\psi\right)\right|\right\}\leq\frac{32}{MK^3}.
		\end{align}
		
		
		\subsection{Overall delay estimation error}
		The overall probability of failing to bound the delay estimation error by $|\epsilon_k|\leq \rho(\log K)^{-2}$ according to \eqref{eq:Tk} thus can be bounded as
		\begin{align}
		P_e&\leq P_{e_1} + P_{e_2}\\
		&\leq \frac{28+128{\Ts }/(\rho M)}{K^2}.
		\end{align}
		Hence the proof.

\end{document}